\documentclass[12pt]{article} 
\usepackage{epsfig}

\renewcommand{\thefootnote}{\fnsymbol{footnote}}

\newcommand{\hard}{^{\mbox{\scriptsize hard}}}
\newcommand{\soft}{^{\mbox{\scriptsize soft}}}
\newcommand{\hql}{\mbox{\scriptsize HQL}}

\begin{document}

\begin{titlepage}
\begin{flushright}
\begin{tabular}{l}
IPPP/02/72 \\
DCPT/02/144
\end{tabular}
\end{flushright}
\vskip0.5cm
\begin{center}
   {\Large \bf $B\to \gamma e\nu$ Transitions from 
    QCD Sum Rules \vspace{0.2cm}\\ on   the Light-Cone}
    \vskip1.3cm {\sc
Patricia Ball\footnote{Patricia.Ball@durham.ac.uk} and 
    Emi Kou\footnote{Emi.Kou@durham.ac.uk}}
  \vskip0.2cm
	IPPP, Department of Physics, 
University of Durham, Durham DH1 3LE, UK\\ 
  \vskip2cm


\vskip2cm

{\large\bf Abstract:\\[10pt]} \parbox[t]{\textwidth}{ 
$B\to\gamma e\nu$ transitions have recently been studied in the
  framework of QCD factorization. The attractiveness of this
    channel for such an analysis lies in the fact that, 
    at least in the heavy quark limit,
    the only hadron involved is the $B$
    meson itself, so one expects a very simple description of the form
    factor in terms of a convolution of the $B$ meson distribution
    amplitude with a perturbative kernel. This description,
    however, does not include contributions suppressed by powers of
    the $b$ quark mass. In this letter, we calculate 
    corrections to the factorized expression which
    are induced by the ``soft'' hadronic component of the photon.
    We demonstrate that the power-suppression of these
    terms is numerically not effective 
    for physical values of the $b$ quark mass and
    that they increase the form factor by about 30\% 
    at zero momentum transfer.
    We also derive a sum rule for $\lambda_B$, the first negative
    moment of the $B$ meson distribution amplitude, and find
    $\lambda_B = 0.6\,$GeV (to leading order in QCD).}
 \vfill
{\em Submitted to JHEP}
\end{center}
\end{titlepage}


\renewcommand{\thefootnote}{\arabic{footnote}}
\addtocounter{footnote}{-2}
\section{Introduction}

With the $B$ factories BaBar and Belle running
``full steam'', $B$ physics has entered the era of precision
measurements. The quality and precision of experimental data
calls for a corresponding 
match in theoretical precision, in particular with regard to
the analysis of nonleptonic decays. Notwithstanding the
fact that a complete solution of the problem appears as elusive as
ever, progress has been made in the description of nonleptonic $B$
decays, in the
heavy quark limit, which were shown to be amenable to perturbative QCD 
(pQCD) factorization
\cite{BBNS1,BBNS2}. In this framework, $B$ decay amplitudes are
 decomposed into a ``soft'' part, for instance a weak decay form factor,
or, in general, 
an intrinsically nonperturbative quantity that evades further
breakdown into factorizable components, and a ``hard'' part which can
be neatly described in a factorized form in terms of a 
convolution of a hard perturbative
kernel, depending only on collinear momenta, with one or more hadron
distribution amplitudes that describe the (collinear) momentum
distribution of partons inside the hadron. Factorization was shown to
hold, for certain $B$ decay channels, to all orders in perturbation theory
\cite{softcoll}, but to date could not be extended to include
contributions that are suppressed by powers of the $b$ quark
mass. For the important channel $B\to K\pi$, in particular, it was
found that factorization breaks down for one
specific class of power-suppressed corrections which are expected to
be numerically relevant \cite{BBNS2}.
Reliable alternative methods for calculating QCD effects in
nonleptonic $B$ decays are
scarce, and only little is known about the generic size of
power-suppressed corrections for instance in $B\to\pi\pi$ \cite{alex}, 
one of the ``benchmark'' channels for measuring the angle $\alpha$ of
the CKM unitarity triangle. There is, however, one channel that can be
treated both in pQCD factorization and by an alternative method,
QCD sum rules: the $B\to\gamma \ell\nu_\ell$
transition. It has been shown recently
\cite{CS,LPW} that $B\to\gamma$ is indeed accessible to collinear
factorization, in contrast to previous findings \cite{Kor} which
indicated the need to include also transverse degrees of freedom in
the convolution. On the other hand, the $B\to\gamma$ transition
can also been investigated in the framework of QCD sum
rules on the light-cone \cite{KSW,AB}. The crucial point here is that
the photon is not treated as an exactly pointlike object with 
standard EM couplings, but that, in addition to that ``hard''
component, it also features a ``soft'' hadronic component whose
contribution to the decay amplitude must not be neglected. 
The ``soft'' component is related to the probability of a
real photon to dissociate, at small transverse separation, into partons
and resembles in many ways 
a (massless) transversely polarized vector meson; like it, it
can be described by
 a Fock-state expansion in terms of distribution amplitudes of
increasing twist. An
analysis of these distribution amplitudes, including terms up to twist-4, has
recently been completed \cite{BBK} and comes in handy for an update
and extension of the previous QCD sum rules analyses of
$B\to\gamma$. Such a reanalysis is the subject of this letter and we 
include in particular one-loop
radiative corrections to the contribution of leading twist-2
distribution amplitude to $B\to\gamma$. In the framework of pQCD
factorization,  
the ``soft'' component of the photon
leads to formally power-suppressed contributions and is hence
neglected in Refs.~\cite{CS,LPW,Kor}. As we shall show in this letter,
this suppression is
not effective numerically due to the large value of the matrix-element
governing its
strength: the magnetic susceptibility of the quark condensate. Our
findings indicate that such contributions are likely to be nonneglible
also in other channels involving photon emission, notably $B\to
K^*\gamma$ and $B\to \rho\gamma$, which are treated in
\cite{BFS,Bosch}. We also compare the QCD sum rule for the hard
part of the $B\to\gamma$ amplitude with the pQCD result and derive a
sum rule for $\lambda_B$, the first negative moment of the $B$ meson
distribution amplitude.


\section{Definition of Relevant Quantities and Outline of 
Calculation}\label{sec2}

First of all, let us define the form factors that describe the
$B\to\gamma$ transition:
\begin{equation}\label{eq1}
\frac{1}{\sqrt{4\pi\alpha}}\,
\langle \gamma(\epsilon^*,q)|\bar{u}\gamma_{\mu}(1-\gamma_5)b
|B^-(p_B)\rangle =\\
-F_V\,\epsilon_{\mu\nu\rho\sigma}\epsilon^{*\nu}v^{\rho}q^{\sigma}
+iF_A[\epsilon^*_{\mu}(v\cdot q)-q_{\mu}(\epsilon^*\cdot v)],
\end{equation}
where $\epsilon^{*}$ and $q$ are the polarisation and momentum vector
of the photon, respectively, and $v=p_B/m_B$ is the four-velocity of
the $B$ meson. The definition of the form factors $F_{A,V}$ in
(\ref{eq1}) is exactly the same as in Ref.~\cite{CS}.\footnote{The
  difference in sign in the 1st term on the r.h.s.\ is due to our
  conventions for the epsilon tensor: we define 
Tr$[\gamma_{\mu}\gamma_{\nu}\gamma_{\rho}\gamma_{\sigma}\gamma_5]=
4i\epsilon_{\mu\nu\rho\sigma}$, in contrast to \cite{CS},  where a
different sign was chosen.}
The form factors depend on $p^2=(p_B-q)^2$ or, equivalently, on the
photon energy $E_\gamma = (m_B^2-p^2)/(2m_B)$. $p^2$ varies in the
physical region $0\leq p^2\leq m_B^2$, which corresponds to $0\leq
E_\gamma\leq m_B/2$. 

The starting point for the calculation of the form factors from QCD
sum rules is the 
correlation function
\begin{eqnarray}\label{eq:corr}
\Pi_{\mu}(p,q)&=&i\int d^4x e^{ipx}
\frac{1}{\sqrt{4\pi\alpha}}\langle\gamma(\epsilon^*,q)|
T\{\bar{u}(x)\gamma_{\mu}(1-\gamma_5)b(x)\bar{b}(0)i\gamma_5u(0)\}
|0\rangle \nonumber \\
&=&-\Pi_{V}\epsilon_{\mu\nu\rho\sigma}\epsilon^{*\nu}p^{\rho}q^{\sigma}
+i\Pi_{A}[\epsilon^*_{\mu}(p\cdot q) - q_{\mu}(\epsilon^*\cdot p)]+\dots
\end{eqnarray}
The dots refer to contact terms which appear for pointlike photons and
for a discussion of which we refer to
Ref.~\cite{contact}; the treatment of the soft component of the photon
involves nonlocal operators, as we shall discuss below, 
and gauge-invariance of $\Pi_\mu$ is
realized explicitly, without contact terms,
by working in the background field method.
 
The method of QCD sum rules \cite{SVZ} exploits the fact that the
correlation function contains information on the form factors in
question: expressing $\Pi_{V(A)}$ via a dispersion relation, one has
\begin{equation}\label{disper}
\Pi_{V(A)}=\frac{f_Bm_BF_{V(A)}}{m_b(m_B^2-p_B^2)}
+\int^{\infty}_{s_0}\,\frac{ds}{s-p_B^2}\,\rho_{V(A)}(s,p^2),
\end{equation}
where the first term on the r.h.s.\ is the contribution of the ground state
$B$ meson to the correlation function, featuring the form factors we
want to calculate, 
and the second term includes all other states coupling to the
pseudoscalar current $\bar{b}i\gamma_5u$, above the 
threshold $s_0$. $f_B$ is the decay constant of the $B$ and defined as 
$\langle B|\bar{b}i\gamma_5u|0\rangle =f_Bm_B^2/m_b$.
The form factors $F_{A,V}$ are obtained, in principle, by equating the
dispersion relation (\ref{disper}) to $\Pi_{A,V}$ calculated for
Euclidean $p_B^2$ by means of an operator product expansion. The sum
over higher states, the 2nd term on the r.h.s.\ of (\ref{disper}), is
evaluated using quark-hadron duality, which means that the hadronic
spectral density $\rho_{V(A)}$ is replaced by its perturbative
equivalent. In order to reduce the model-dependence associated with
that procedure, one  subjects the whole expression to a
Borel-transformation, which results in an exponential suppression of
the continuum of higher states:
\begin{equation}
B_{M^2}\Pi_{V(A)} = \frac{f_Bm_BF_{V(A)}}{m_b}\,e^{-m_B^2/M^2} +
\int_{s_0}^\infty ds\,\rho^{\mbox{\scriptsize pert}}_{V(A)}(s) e^{-s/M^2}.
\label{SRs}
\end{equation}
The relevant parameters of the sum rule are then $M^2$, the Borel
parameter, and $s_0$, the continuum threshold, and our results for 
the form factors will depend (moderately) on these parameters.

\begin{figure}
\centerline{\epsffile{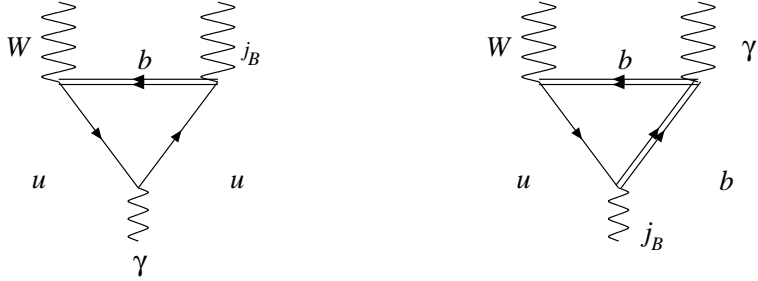}}
\caption[]{\small Diagrams with hard photon emission.}\label{fig:1}
\end{figure}

As already mentioned in the introduction, the 
photon does not only have EM pointlike couplings to the quarks, but
also soft nonlocal ones, and we write $F_{A,V} =
F_{A,V}\hard + F_{A,V}\soft$. The separation between
these two components is of course scheme-dependent and we define them
in the $\overline{\rm MS}$ scheme.
For the pointlike component of the photon, $\Pi_\mu$ is just a
three-point correlation function, described by the triangle diagrams
shown in Fig.$\,$\ref{fig:1}. These diagrams, and also the leading
nonperturbative correction which is
proportional to the quark condensate, have been calculated in
Refs.~\cite{KSW,AB}, and we confirm the results; we will come back to
the hard contributions in Sec.~\ref{sec:5}. The calculation of
radiative corrections to the three-point function, although highly
desirable, is beyond the scope of this letter 
whose main emphasis is on calculating 
the soft photon contributions which become relevant for a certain
configuration of virtualities, namely
$m_b^2-p_B^2\geq O(\Lambda_{\rm QCD}m_b)$ and $m_b^2-p^2\geq
O(\Lambda_{\rm QCD}m_b)$, i.e.\ $E_\gamma\gg \Lambda_{\rm QCD}$.  
In this regime, the 
integral in Eq.$\,$(\ref{disper}) is dominated by light-like distances
and can be expanded around the light-cone:
\begin{equation}\label{eq:3}
\Pi_{V(A)}(p_B^2,E_\gamma) = Q_u \chi(\mu_F) \langle\bar u u\rangle(\mu_F)
\sum_n \int_0^1 du\, \phi^{(n)}(u;\mu_{F})
T_{V(A)}^{(n)}(u;p_B^2,E_\gamma;\mu_{F}).
\end{equation}
This is a factorization formula expressing the
correlation function as a convolution of genuinely nonperturbative and
universal  distribution amplitudes (DA) $\phi^{(n)}$ with 
process-dependent hard kernels $T^{(n)}$, to be calculated in perturbation
theory; the overall factor $Q_u \chi \langle\bar u u\rangle$
($Q_u=2/3$) is pulled out for later convenience. 
In the above equation, $n$ labels the twist of operators and
$\mu_{F}$ is the factorization scale. The restriction
on $E_\gamma$
implies that $F\soft_{A,V}$ cannot be calculated for all photon energies; to be
specific, we restrict ourselves to $E_\gamma>1\,$GeV.
The factorization formula holds if the infrared divergencies occurring
in the calculation of $T^{(n)}$ are such that they can be absorbed into
the universal distribution amplitudes $\phi^{(n)}$ and if the convolution
integral converges. We find that this is indeed the case,
at least for the leading twist-2 contribution and to first order in
$\alpha_s$. We also would like to stress that the above factorization
formula has got nothing to do with the pQCD factorized expression for
$F_{A,V}\hard$ derived in \cite{CS,LPW} (although we will discuss the
relevance of our findings as compared to theirs in Sec.~4) and that 
it is valid for arbitrary values
of the $b$ quark mass --- there is
no need to restrict oneself to the heavy quark limit.

Let us now define the photon DAs that enter Eq.$\,$(\ref{eq:3}). As
mentioned above, we shall work in the background field gauge, which
means that the expression (\ref{disper}), formulated for an outgoing
photon, gets replaced by a correlation function of the time-ordered
product of two currents in the vacuum, which is populated by an arbitrary EM
field configuration $F_{\alpha\beta}$,
\begin{equation}
\Pi_F^{\mu}(p,q)=i\int d^4x e^{ipx}
\frac{1}{\sqrt{4\pi\alpha}}\langle 0|
T\{\bar{u}(x)\gamma^{\mu}(1-\gamma_5)b(x)\bar{b}(0)i\gamma_5u(0)\}
|0\rangle_F,
\end{equation}
where the subscript $F$ indicates that an EM background field $B_\mu$
is included in the action. The calculation is explicitly
gauge-invariant, and it is only in the final step that we select one
specific field configuration corresponding to an outgoing photon with
momentum $q$:
$$F_{\alpha\beta}(x)\to -i (\epsilon^*_\alpha q_\beta -
\epsilon^*_\beta q_\alpha) e^{iqx}.$$
Following Ref.~\cite{BBK}, 
we define the leading-twist DA
as the vacuum expectation value of the nonlocal quark-antiquark
operator with light-like separations, in the EM background field
configuration $F_{\alpha\beta}$:
\begin{equation}
   \langle 0|\bar q(z)\sigma_{\alpha\beta}[z,-z]_F q(-z)  |0\rangle_F =
    e_q\, \chi\, \langle \bar q q\rangle
 \int\limits_0^1 \!du\, F_{\alpha\beta}(-\xi z) \phi_\gamma(u)\,,
\label{T2}
\end{equation}
where $z^2=0$ and $[z,-z]_F$ is the path-ordered gauge-factor
\begin{equation}
[z,-z]_F ={\rm P}\!\exp\left\{i\!\!\int_0^1\!\! dt\,2z_\mu
  [g A^\mu((2t-1)z)+e_q B^\mu((2t-1)z)]\right\},
\end{equation}
including both the gluon field $A_\mu=A_\mu^a \lambda^a/2$ and the EM
background field $B_\mu$. $e_q$ is the electric charge of the light
quark $q$, e.g.\ $e_u=2/3 \sqrt{4\pi\alpha}$, 
$\langle \bar q q\rangle$ the quark
condensate and $\chi$ its magnetic susceptibility -- defined via the
local matrix element $\langle 0|\bar q\sigma_{\alpha\beta} q |0\rangle_F =
    \,e_q\, \chi\, \langle \bar q q\rangle \,F_{\alpha\beta}$, which
    implies the normalization $\int_0^1 du\phi_\gamma(u)=1$.
$\phi_\gamma$ can be expanded in terms of contributions of increasing
    conformal spin (cf.\ \cite{BBKT} for a  detailed discussion of the
    conformal expansion of DAs):
\begin{equation}
\phi_\gamma(u,\mu)  = 6u(1-u)
  \left[1+ \sum_{n=2,4,\ldots}^\infty \phi_n(\mu)
  C^{3/2}_n(2u-1)\right],
\label{T2DA}
\end{equation}
where $C_n^{3/2}$ are Gegenbauer polynomials and $0\leq u\leq 1$ is
the collinear momentum fraction carried by the quark. The usefulness
of the conformal expansion lies in the fact that the Gegenbauer moments
$\phi_n(\mu)$ renormalize multiplicatively in LO perturbation theory.
At higher twist, there exists a full plethora of photon DAs, which we
refrain from defining in full detail, but refer the reader to the discussion in
Ref.~\cite{BBK}, in particular Sec.~4.1.

\begin{figure}
\centerline{\epsffile{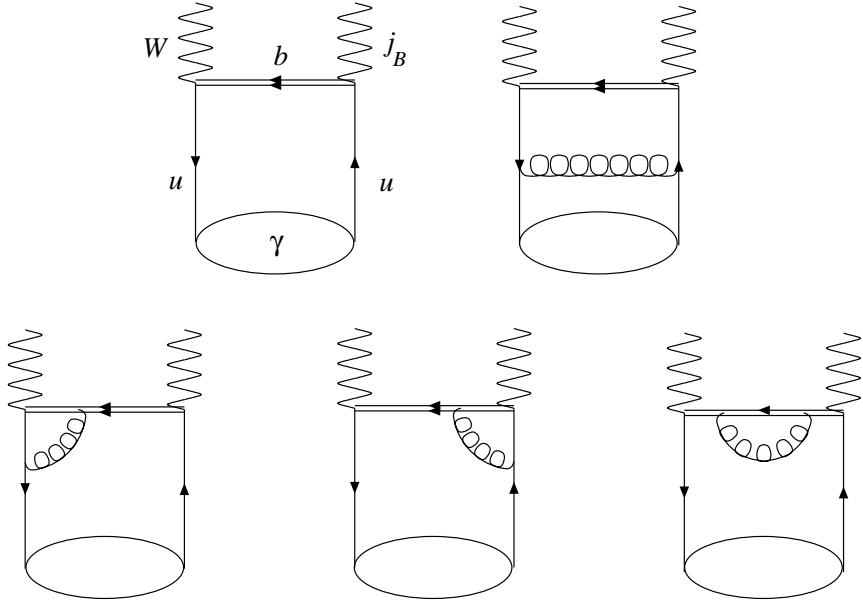}}
\caption[]{\small Leading twist-2 soft photon contributions to the
  correlation function (\ref{eq:corr}).}\label{fig:2}
\end{figure}

The task is now to calculate the hard kernels $T^{(n)}$, to $O(\alpha_s)$ in
leading-twist and tree-level for higher twist. As for the former, the
relevant diagrams are shown in Fig.~\ref{fig:2}.
The diagrams are IR divergent, which divergence can be absorbed into
the DA $\chi\langle\bar u u \rangle \phi_\gamma$; we have checked that
this is indeed the case. A critical point in pQCD calculations
involving heavy particles is the possibility of soft divergencies,
which manifest themselves as divergence of the $u$ integration at the
endpoints; the success of the approach advocated in
\cite{BBNS1,BBNS2} relies precisely on the fact that it yields 
a factorization formula for
nonleptonic decays where these divergencies are absent
in the heavy quark limit (but come back at order $1/m_b$). 
We find that there are no soft divergencies in our case\footnote{Note
  that this statement applies to the {\em full} correlation function
  with finite $m_b$,
  not only to the heavy quark limit.}, which reiterates 
what has been found for 
other heavy-light transitions, cf.\ \cite{BBB,BB98,BZ}.
We thus confirm the factorization formula (\ref{eq:3}) to
$O(\alpha_s)$ at twist-2. Explicit expressions for the spectral
densities of the diagrams are too bulky to be given
here; they can be obtained from the authors. 
At this point we only note that, to
leading-twist accuracy, $\Pi_A \equiv \Pi_V$.
 
As for the higher-twist contributions, we obtain the following results:
\begin{eqnarray}
\Pi_A&=&
-Q_uf_{3\gamma}\int^1_0du \frac{\bar{\psi}^{(v)}(u)}{s^2} m_b 
+\frac{1}{2}Q_u\langle \bar{u}u\rangle 
\int^1_0du \frac{\bar{h}(u)}{s^2}\left[1+\frac{2m^2}{s}\right] \\
&&-Q_u\langle \bar{u}u\rangle \int^1_0dv\int\mathcal{D}\underline{\alpha}
\frac{S(\underline{\alpha})}{\tilde{s}^2}(1-2v) 
+\frac{1}{6}Q_u\langle \bar{u}u\rangle \int^1_0dv\int\mathcal{D}
\underline{\alpha}
\frac{\tilde{S}(\underline{\alpha})}{\tilde{s}^2} \nonumber \\
&&-2Q_u\langle \bar{u}u\rangle \int^1_0dv\int\mathcal{D}\underline{\alpha}
\frac{p\cdot q}{\tilde{s}^3}
\left[\bar{T}_1(\underline{\alpha})-(1-2v)\bar{T}_2
(\underline{\alpha})+(1-2v)\bar{T}_3(\underline{\alpha})-\bar{T}_4
(\underline{\alpha})\right], \nonumber \\ 
\Pi_V &=&\frac{1}{2}Q_uf_{3\gamma}\int^1_0du \frac{\psi^{(a)}(u)}{s^2} m_b \\
&&-Q_u\langle \bar{u}u\rangle \int^1_0dv\int\mathcal{D}\underline{\alpha}
\frac{S(\underline{\alpha})}{\tilde{s}^2} 
+\frac{1}{6}Q_u\langle \bar{u}u\rangle \int^1_0dv\int\mathcal{D}
\underline{\alpha}
\frac{\tilde{S}(\underline{\alpha})}{\tilde{s}^2}(1-2v) \nonumber \\
&&-2Q_u\langle \bar{u}u\rangle \int^1_0dv\int\mathcal{D}\underline{\alpha}
\frac{p\cdot q}{\tilde{s}^3}\left[\bar{T}_1(\underline{\alpha})-\bar{T}_2
(\underline{\alpha})+(1-2v)\bar{T}_3(\underline{\alpha})-(1-2v)\bar{T}_4
(\underline{\alpha})\right],\nonumber
\end{eqnarray}
where $s=m_b^2-(p+uq)^2$ and $\tilde{s}=m_b^2-(p+\zeta q)^2$ with  
$\zeta=(\alpha_{\bar{q}}-\alpha_q+(2v-1)\alpha_g+1)/2$. 
Definitions and explicit expressions for the higher twist DAs  
$\bar{\psi}^{(v)}(u)$ and $\psi^{(a)}(u)$ (two-particle twist-3),  
$\bar{h}(u)$ (two-particle twist-4),   
$S(\underline{\alpha })$, $\tilde{S}(\underline{\alpha })$ (three-particle 
twist-3) and $\bar{T}_i(\underline{\alpha })$ (three-particle twist-4) 
can be found in \cite{BBK}. 
Note that in contrast to the leading-twist results,
$\Pi_V^{\mbox{\scriptsize higher twist}} \neq 
\Pi_A^{\mbox{\scriptsize higher twist}}$.

\section{\boldmath Numerical Results for $F_{A,V}\soft$}

Before presenting numerical values for the soft part of the
form factors, we first discuss
the numerical input to the sum rule (\ref{SRs}). As for the
photon DA, we use the values and parametrizations derived in
\cite{BBK}, notably for the normalization of the twist-2 matrix
element (\ref{T2}):
$$
(\chi\langle\bar u u \rangle)(1\,\mbox{GeV}) = -(0.050\pm
  0.015)\,\mbox{GeV}.
$$
As discussed in 
\cite{BBK}, there is no conclusive evidence for $\phi_\gamma$ to
differ significantly from its asymptotic form, so we set 
$$
\phi_\gamma(u) = 6 u (1-u).
$$
The remaining hadronic matrix elements characterizing higher-twist DAs
are detailed in \cite{BBK}. Note that we evaluate scale-dependent
quantities at the factorization scale $\mu_F^2 = m_B^2 - m_b^2$
\cite{bel}; the dependence of the form factors on $\mu_F$ is very
small, as all numerically sizeable contributions are now available to
NLO in QCD, which ensures good cancellation of the residual scale
dependence.

As for the remaining parameters occurring in (\ref{SRs}), we have
the sum rule specific parameters $M^2$ and $s_0$, that is the
Borel parameter and the
continuum threshold, respectively. 
In addition, the sum rule depends on $m_b$, the b quark mass, and $f_B$, the
leptonic decay constant of the $B$. $f_B$ can in principle be
measured from the decay $B\to \ell\bar\nu_\ell$, which, due to the expected
smallness of its branching ratio, BR$\,\sim O(10^{-6})$,
has, up to now, escaped experimental detection. $f_B$ is one of the
best-studied observables in lattice-simulations with heavy quarks; the
current world-average from unquenched calculations with two dynamical
quarks is $f_B = (200\pm 30)\,$MeV \cite{fBlatt}. It can also be
calculated from QCD sum rules: the most recent determinations
\cite{fBSR} include $O(\alpha_s^2)$ corrections and find $(206\pm 20)\,$MeV
and $(197\pm 23)\,$MeV, respectively. For consistency, we do not use
these results, but replace $f_B$ in (\ref{SRs}) by its QCD sum rule
to $O(\alpha_s)$ accuracy, including the dependence on $s_0$ and $M^2$,
and use the corresponding ``optimum'' ranges of continuum threshold and
Borel parameter also in evaluating the Borel-transformed correlation
function $\Pi_{V(A)}$, i.e.\ the l.h.s.\ of (\ref{SRs}). 
For the $b$ quark mass, we use an average over
recent determinations of the $\overline{\rm MS}$ mass,
$\overline{m}_{b,\overline{\rm MS}}(\overline{m}_b) = (4.22\pm
0.08)\,$GeV \cite{bquark,latmasses},
which corresponds to the one-loop pole-mass
$m_{b,{\rm 1L-pole}}=(4.60\pm 0.09)\,$GeV. With these values we find 
$f_B =(192\pm 22)\,$GeV (the error only includes variation with $m_b$
and $M^2$, at optimized $s_0$), in very good agreement with both
lattice and QCD sum rules to $O(\alpha_s^2)$ accuracy. For $m_b = 
(4.51,4.60,4.69)\,$GeV the optimized $s_0$ are
$(34.5,34.0,33.5)\,$GeV$^2$, and the relevant range of $M^2$ is
(4.5--8)$\,$GeV$^2$.

\begin{figure}
$$\epsffile{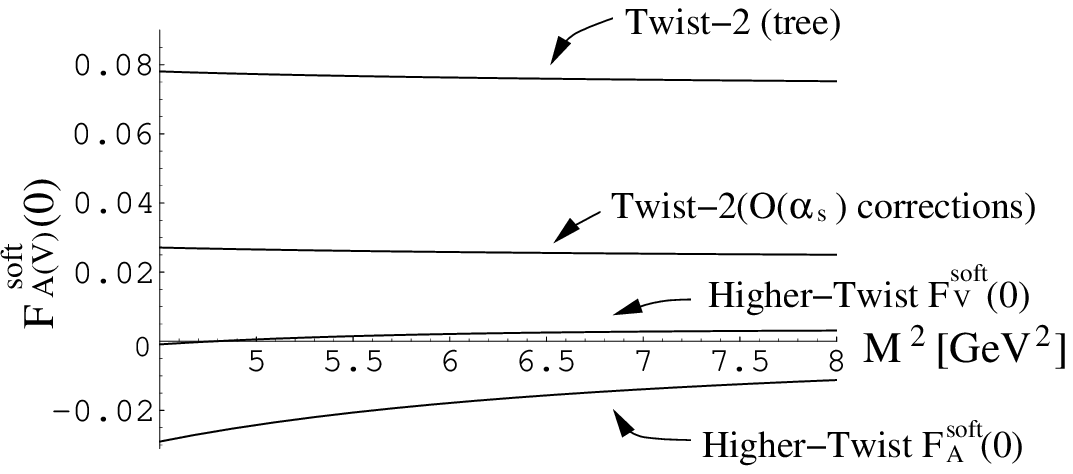}$$
\vskip-0.5cm
\caption[]{{\small The soft photon contributions to $F_{V(A)}(0)$
in dependence of the Borel parameter $M^2$. The twist-2 contributions
to both form factors are equal. The plot refers to  
$m_b=4.6\,$GeV and $s_0=34 \,$GeV$^2$.   }}
\label{figure2}
$$\epsffile{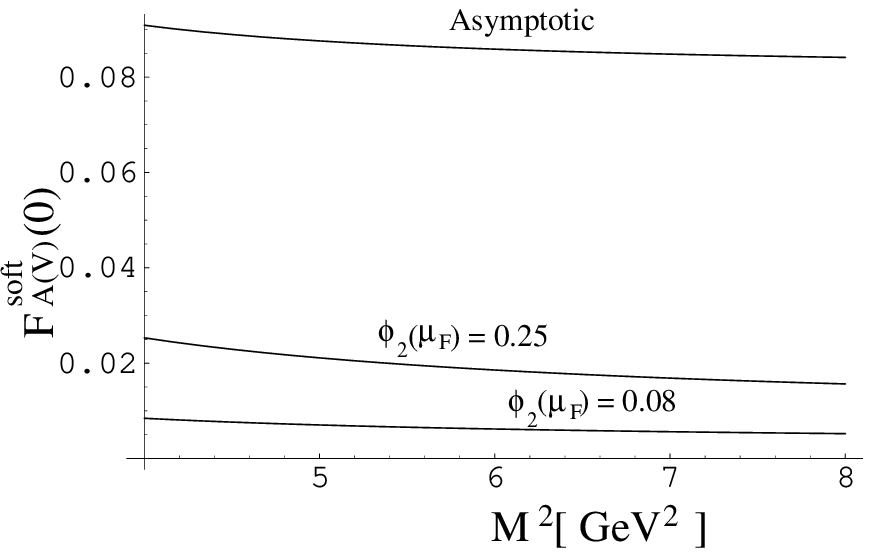}$$
\vskip-1cm
\caption[]{{\small Contributions to $F\soft_{V(A)}$ 
from different values $\phi_2$ of the 2nd Gegenbauer moment of the
photon DA ($m_b=4.6\,{\rm GeV},\,s_0 = 34\,{\rm GeV}^2$).  }}
\label{figure4}
$$\epsfxsize=\textwidth\epsffile{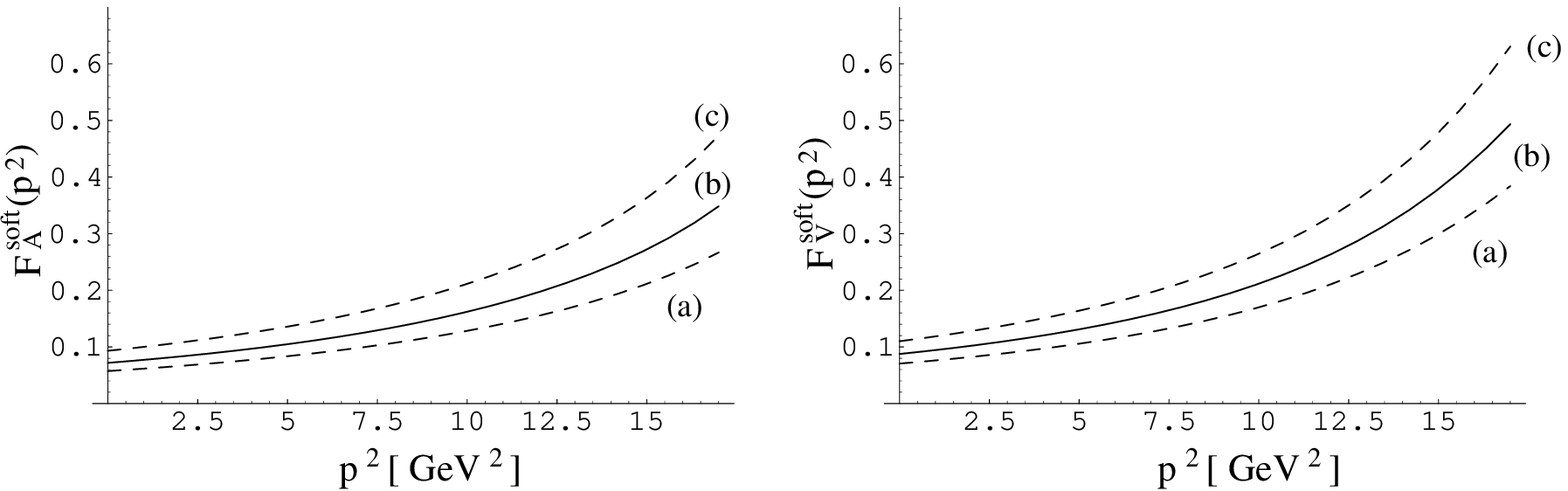}$$
\vskip-0.5cm
\caption[]{{\small Dependence of $F\soft_{V(A)}$ on the momentum
    transfer $p^2$. Solid line: light-cone sum rule for central values
    of input parameters. Dashed lines: spread of $F\soft_{V(A)}$
upon variation of input parameters. The exact parameter sets for the
    curves (a) to (c) are listed in Tab.~{\ref{table1}}. 
Note that the uncertainty originating from $M^2$ and $s_0$ is less than 
$10\%$.}}
\label{figure3}
\end{figure}

In Fig.~\ref{figure2} we plot the different contributions to
$F_{V,A}\soft(0)$ as function of $M^2$, for $m_b=4.6\,$GeV, $s_0 =
34\,$GeV$^2$ and the central value of $\chi\langle\bar u u\rangle$. It
is evident that the sum rule is dominated by twist-2 contributions and
that both radiative corrections and higher-twist terms are well under
control. Note also the minimal sensitivity to $M^2$ which indicates a
``well-behaved'' sum rule. Varying $m_b$, $s_0$ and the other input
parameters within the ranges specified above, we find
\begin{equation}\label{007}
F\soft_A(0) = 0.07\pm 0.02, \qquad F_V\soft(0) = 0.09\pm 0.02.
\end{equation}
As mentioned before, the above results are obtained using the
asymptotic form of the twist-2 photon DA. Although there is presently
no evidence for nonzero values of
higher Gegenbauer moments, the $\phi_n$ in  (\ref{T2DA}), it may be
illustrative to estimate their possible impact on the form factors. As
a guideline for numerics, we choose $\phi^\gamma_2$ to be equal to
$\phi_2^{\rho_\perp}$, its analogue for the 
transversely polarized $\rho$ meson, as determined in \cite{BB96}:
$\phi^{\rho_\perp}_2(1\,{\rm GeV}) = 0.2\pm 0.1$. In
Fig.~\ref{figure4} we plot the twist-2 
contribution to the form factors obtained
with the asymptotic $\phi_\gamma$ and the corrections induced by
nonzero $\phi_2^\gamma(\mu_F)$. It is clear that the effect is moderate
and at most about 20\%. It is also interesting to note that positive values
of $\phi_2^\gamma$, which are in accordance with assuming $\rho$ meson
dominance for the photon, increase the  form factor,
which means that our results are likely to be an {\em underestimate}
of the soft contributions rather than the contrary.

In Fig. \ref{figure3}, we show the dependence of $F_{A,V}\soft(p^2)$
on the momentum transfer $p^2$, including the variation of all
input parameters in their respective ranges. The form factors can be
fitted by the following formula:
\begin{equation}
F_{A(V)}\soft(p^2)=\frac{F_{A(V)}\soft(0)}
{1-a_{A{(V)}}\left(p^2/m_B^2\right)+b_{A(V)}
\left(p^2/m_B^2\right)^2}\,.
\label{interpolation}
\end{equation}
The fit parameters $F\soft_{A(V)}(0)$, $a_{A(V)}$ and $b_{A(V)}$  for each 
curve in the figure are given in Tab.~\ref{table1}. The above formula
fits the full sum rule results to  1\% accuracy for $0<p^2<17\,{\rm
  GeV}^2$, which corresponds to $1\,{\rm GeV}< E_\gamma< m_B/2$. 
Note that the uncertainty induced by $M^2$ and $s_0$ is very small: 
up to $5\% $ for $F_V$ and up to $10\% $ 
for $F_A$. The main theoretical uncertainty comes from 
$\chi\langle\bar{q}q\rangle$.

\begin{table}
\begin{center}
\begin{tabular}{|c|c|c|c|}\hline
Input parameters                       & (a)    &    (b)   &   (c)   \\ \hline
$m_b$ [GeV]             & 4.69   & 4.60     & 4.51    \\
$s_0\ \mbox{[GeV}^2]$  & 33.5   & 34.0     & 34.5    \\
$M^2\ \mbox{[GeV}^2]$  & 8      & 6        & 5       \\
$\chi\langle\bar{u}u\rangle(\mu=1\,$GeV) [GeV]
                        & $-0.035$  & $-0.050$     & $-0.065$ \\ \hline
Fit parameters   & (a)    &    (b)   &   (c) \\ \hline
$F_A\soft(0)$                & 0.057  & 0.072    & 0.093    \\
$a_A$                   & 1.97   & 1.97     & 1.95     \\
$b_A$			& 1.18   & 1.15     & 1.08     \\ \hline
$F_V\soft(0)$                & 0.071  & 0.088    & 0.110    \\
$a_V$                   & 2.09   & 2.08     & 2.05     \\
$b_V$			& 1.25   & 1.22     & 1.16     \\ \hline
\end{tabular}
\end{center}
\caption[]{{\small 
Input parameter sets for Fig.~\ref{figure3} and  
fit parameters for Eq.~(\ref{interpolation}). }}
\label{table1}
\end{table}

\section{\boldmath Calculation of
  $\lambda_B$ -- Comparison to pQCD}\label{sec:5} 

As mentioned above, the 
$B\to\gamma$ form factors have also been calculated in the
framework of pQCD factorization \cite{Kor,CS,LPW}. This
approach employs the limit $m_b\to\infty$ (HQL), in which the soft
contributions vanish. The form factors $F\hard_{A(V),\hql}$ are
equal and at tree level given by
\begin{equation}\label{FApQCD}
F_{A,\hql}\hard(E_\gamma) \equiv F_{V,\hql}\hard(E_\gamma)
 = \frac{f_B m_B Q_u}{2\sqrt{2} E_\gamma} 
\int\limits_0^\infty dk_+\,\frac{\Phi_+^B(k_+)}{k_+} =: \frac{f_B m_B
  Q_u}{2 E_\gamma}\,\frac{1}{\lambda_B},
\end{equation}
where the light-cone DA $\Phi_+^B$ of the $B$ meson depends on the
momentum of the light spectator quark, $k_+$.
The calculation of radiative corrections to this formula has been the
subject of a certain controversy, cf.\ Refs.\ \cite{Kor,CS,LPW}.
The parameter $\lambda_B$ does not scale with
$m_b$ in the HQL and hence is of natural size
$O(\Lambda_{\mbox{\scriptsize QCD}})$; it has been quoted as
$\lambda_B = 0.3\,$GeV \cite{BBNS1} and $\lambda_B = (0.35\pm
0.15)\,$GeV \cite{CS}, but without calculation. Although presently any
statement about the numerical size of $\lambda_B$ appears slightly
precarious, since $\Phi_B^+$ and hence $\lambda_B$ depend in a yet unknown
way on the factorization scale $\mu_F$, we nonetheless venture to
present the (to the best of our knowledge) first calculation of
$\lambda_B$. To that purpose, we recall that the hard contribution to
$F_{A,V}$ can be obtained, in the QCD sum rule approach, from the local
contributions to the correlation function $\Pi_{A,V}$,
Eq.~(\ref{eq:corr}), which, to leading order in perturbation theory,
correspond to the diagrams shown in Fig.~\ref{fig:1} and have been
calculated in \cite{KSW}. In order to extract $\lambda_B$ via
(\ref{FApQCD}) from the local QCD sum rule for $F\hard_{A,V}$, we
first have to find its HQL. QCD sum rules
in the heavy quark limit have actually been studied in quite some
detail, cf.~\cite{BBBD}, with the following result for the scaling
relations of the sum rule specific parameters $M^2$ and $s_0$:
\begin{equation}\label{scaling}
M^2\to 2 m_b\tau,\quad s_0\to m_b^2 + 2 m_b \omega_0.
\end{equation}
Applying these relations to the sum rule for $F\hard_{V,A}$
and using (\ref{FApQCD}), we 
obtain the following expression:
\begin{equation}\label{lamma}
e^{-\bar\Lambda/\tau}\,\frac{f_B^2
  m_B^2}{m_bE_\gamma}\,\frac{1}{\lambda_B} = \frac{3}{\pi^2 E_\gamma}
  \int_0^{\omega_0} d\omega \,\omega \, e^{-\omega/\tau}.
\end{equation}
We note that the factor $1/E_\gamma$ on the r.h.s.\ 
arises automatically in the HQL
of the correlation function. $\bar\Lambda$ is the binding energy of
the $b$ quark in the $B$ meson, $\bar\Lambda = m_B - m_b$. On the
l.h.s.\ of (\ref{lamma}), the expression $f_B^2 m_B^2/m_b$ still
contains $1/m_b$ corrections; in the rigorous HQL we have $f_B^2
m_B^2/m_b \to f_{\mbox{\scriptsize stat}}^2$. $f_{\mbox{\scriptsize
    stat}}$ is known both from lattice calculations and QCD sum rules;
in the same spirit that we applied for calculating $F_{A,V}\soft$,
we replace, for the numerical evaluation of $\lambda_B$,
 $f_{\mbox{\scriptsize stat}}^2$ by its sum rule \cite{BBBD}:
\begin{equation}\label{fstat}
f_{\mbox{\scriptsize stat}}^2 e^{-\bar\Lambda/\tau} = \frac{3}{\pi^2}
\int_0^{\omega_0} d\omega\,\omega^2 e^{-\omega/\tau},
\end{equation}
where we have suppressed (small) condensate contributions. 
Combining (\ref{lamma}) and (\ref{fstat}), we find
\begin{equation}\label{lambda}
\lambda_B = \frac{\displaystyle \int_0^{\omega_0} d\omega\,\omega^2
  e^{-\omega/\tau}}{\displaystyle 
\int_0^{\omega_0} d\omega\,\omega e^{-\omega/\tau}}.
\end{equation}
This is our sum rule for $\lambda_B$, which, admittedly, is the first
rather than the last word in the story of how to calculate $\lambda_B$.
It can and should be improved by including both radiative
(which will also settle the issue of scale dependence) and
nonperturbative corrections. For fixing the sum rule parameters
$\omega_0$ and $\tau$, we exploit the fact that
$\bar\Lambda = m_B-m_b \approx 0.7\,$GeV is known; the corresponding
sum rule reads
\begin{equation}
\bar\Lambda = \frac{\displaystyle\int_0^{\omega_0} d\omega\,\omega^3
  e^{-\omega/\tau}}{\displaystyle
\int_0^{\omega_0} d\omega\,\omega^2 e^{-\omega/\tau}},
\end{equation}
which can be derived from (\ref{fstat}) by taking one derivative in
$1/\tau$. None of the sum rules (\ref{lamma}), (\ref{fstat}),
(\ref{lambda}) is ``well-behaved'' in the sense
that they feature no stability plateau in
$\tau$, since the perturbative term is not counterbalanced by a
nonperturbative one. Nonetheless, taking
$0.5\,\mbox{GeV}<\tau<1\,\mbox{GeV}$ as indicated by our preferred
range of $M^2$ used above and the scaling laws (\ref{scaling}), we
find $\omega_0 = 1\,$GeV and then, from (\ref{lambda}),
$$\lambda_B \approx 0.57\,{\rm GeV}.$$
Given the neglect of $O(\alpha_s)$ corrections and 
nonperturbative terms, it is difficult to attribute
an error to that number. For this reason, we check if our result is
compatible with the local duality approximation, which corresponds to
the limit
$\tau\to\infty$. From (\ref{lamma}) and (\ref{fstat}) we then
obtain the interesting relation $$\lambda_B =
\frac{8}{9}\,\bar\Lambda,$$
which for $m_b = 4.6\,$GeV implies $\lambda_B = 0.6\,$GeV. 
We thus conclude that the value of $\lambda_B$ is set by $\bar\Lambda$
rather than $\Lambda_{\mbox{\scriptsize QCD}}$ and  depends strongly on
the actual value of $m_b$; at present, nothing meaningful can be said about the
error associated with $\lambda_B$ and we  thus quote as our final
result
\begin{equation}
\lambda_B = 0.6\,{\rm GeV}.
\end{equation}

We are now in a position to compare the numerical size of the pQCD
result $F\hard_{A(V),\hql}$ to $F_{A(V)}\soft$. At tree level, we have
$F_{A(V),\hql}\hard(E_\gamma=m_B/2)=0.22,$ according to
(\ref{FApQCD}). Notwithstanding the additional hadronic uncertainties
involved at NLO in QCD, we employ the models for $\Phi_B^+$ advocated
in \cite{CS} to obtain $F_{A(V),\hql}^{\mbox{\scriptsize
    hard},NLO}(E_\gamma = m_B/2)=0.21\pm 0.01$ for $\mu^2_F =
m_B^2-m_b^2$, our choice of the factorization scale. 
This number has to be compared
to (\ref{007}), the soft contributions. We conclude that 
$F_{A,V}\soft/F\hard_{A(V),\hql}\approx 0.3$ at maximum photon energy
so that the parametrical scaling $F_{A,V}\soft\sim O(1/m_b)$ is
numerically relaxed.

\section{Summary and Conclusions}

The relevance of $B$ physics for extracting information on weak
interaction parameters and new physics is limited by our lack of 
knowledge on nonperturbative QCD. Recent progress in describing the
notoriously difficult nonleptonic decays in perturbative QCD
factorization has raised hopes that a sufficiently accurate solution
to the problem is around the corner. As much as this is a highly
desirable goal, it» is nonetheless necessary to critically examine the 
theoretical uncertainty of the method, which, at least at present, is
set by the restriction to the heavy quark limit. A direct
{\em theoretical} test of pQCD factorization in nonleptonic decays is
currently not feasible, and any significant {\em experimental}
deviation of measured decay rates or CP asymmetries
from pQCD predictions is as likely to be attributed to new physics
effects as to uncertainties in the predictions themselves. An indirect
theoretical test becomes however possible in the admittedly
phenomenologically not very attractive channel $B\to\gamma e\nu$, where
alternative methods of calculation exist and allow one to assess 
effects suppressed by powers of $m_b$. In this letter, we have
calculated corrections to the hard pQCD form factors, which are
parametrically suppressed by one power of $m_b$. 
These ``soft'' corrections are induced by
photon emission at large distances and involve the hadronic structure
of the photon. 
We have also presented the first calculation of $\lambda_B$, the first
negative moment of the $B$ meson distribution amplitude, a very
relevant parameter for pQCD calculations. The calculation is
admittedly rather crude, but amenable to
improvement. Comparing the numerical size of the  pQCD  result to
the soft contributions, we found that the latter are indeed sizeable,
$O(30\%)$. This result implies an immediate caveat for
pQCD analyses involving photon emission, in particular $B\to
K^*\gamma$ and $B\to\rho\gamma$, e.g.\ \cite{BFS, Bosch}. In a wider
sense, it also adds a possible questionmark to pQCD analyses of 
purely hadronic $B$ decays
and emphasizes the relevance of power-suppressed corrections to the
heavy quark limit.


\bigskip

\noindent {\large\bf Acknowledgements} \\
We are grateful to C. Sachrajda for useful discussions.

\end{document}